# EKSPERIMENTAL EVIDENCE OF ELECTRIFICATION PROCESSES AT THE 2009 L'AQUILA EARTHQUAKE MAINSHOCK

P. Nenovski

University Center of Space Research and Technology, Sofia University, Sofia, Bulgaria

**Abstract**

Two types of coseismic magnetic field events are simultaneously observed: transient offset events and magnetic field signal that occurred at the destructive, Mw6.1 LAquila earthquake (EQ) mainshock. The offset event, conventionally interpreted as a signature of piezomagnetic effects, however could not be ascribed as such. The reason is that the presently known geology of the L'Aquila basin consists mainly of carbonates, dolomites and limestone, thus it does not suggest an appearance of piezomagnetic related effects under EQ fracture/slip events. The second type of coseismic event, the transient magnetic signal starts simultaneously with the offset event and reaches amplitude of 0.8 nT in the total magnetic field. The signal is local one, in the sense that its form differs from the signals of ionospheric/magnetospheric origin confirmed (indirectly) by additional magnetic field data in Italy and Central Europe. The reliability of the observed local signal is examined also: it follows from the fact that the transient signal is recorded by two different measurements: i) in components (by fluxgate magnetometers) and 2) absolute one (overhauser magnetometer), it persists after the seismic wave train passage but within the first five minutes after the EQ shock and is not a consequence of ionosphere disturbances caused by the seismic wave train.
Its amplitude shape resembles diffusion-like form with time scale characteristics that are indicative for a source deep in the crust. The polarity of the transient signal is in the horizontal plane and nearly parallel to the L'Aquila fault strike, i.e. along the NW-SE direction.

**Introduction.**

Crustal deformation and earthquake related fault failure processes are accompanied with electric and magnetic fields generation mechanisms which include piezomagnetism, stress/conductivity and electrokinetic effects, charge generation processes, charge dispersion, induction, remagnetization and demagnetization effects, etc. A complete discussion of physical limitations, constraints, and frequency limitations placed on these processes has been provided (Johnston, 2002). According to Johnson (2002) a careful work still needs to be done to convincingly demonstrate causality between "precursory" EM signals and earthquakes and consistency with complementary geophysical data as the state of stress, strain, material properties, fluid content, and approach to failure of the Earth's crust in seismically active regions.

It is close to mind that if reliable magnetic and electric field observations are indeed EQ source related, clear signals should occur at the time of large local earthquakes mainshocks because the primary energy release occurs at this time. These signals should scale with the earthquake moment (size) and source geometry (Johnson, 2002).

Presently, observations of co-seismic magnetic field signals recorded at the seismic earthquake (EQ) mainshocks are rare and limited mainly to single field offsets and, in extreme cases, electromagnetic wave-like disturbances similar to the seismic waves in form and duration when passing through the observation point. A lack of sufficient observations of anomalous and transient electric and/or magnetic signals at the EQ mainshock seems indeed suspiciously. The reasons for the absence of signals during the earthquake shocks are probably rooted in at least two facts: 1) the most of the magnetic field measurements documented in literature happen at great distances – hundreds even thousands of km. Cases where earthquakes occur in the vicinity of the measurement are rare (Loma Prieta (1989), Northridge (2004), L'Aquila (2009), etc.). It is not strange: it is well known that magnetic field signals cease rapidly with distance; 2) In seeking pre-seismic signals, the conventional approach assumes intervals of low geomagnetic activity and night hours, namely such intervals have been preferred and analyzed. This suggests that a lot of possible and expected co-seismic signals next to and at the earthquake mainshock events (occurred mostly at hundreds km distances to epicenter and/or in periods of strong and moderate geomagnetic activity) might be missed. Therefore, coseismic magnetic field signals, if these exist, need to be detected and analyzed using data from observations occurred close to EQ epicenter and under low geomagnetic activity conditions. The field signatures, as the magnitude and shape of the signal, its spectral components and polarization, however reflect the origin and physical properties of the source, possible mechanisms, material property and geometry. The coseismic electric and magnetic field signatures, if there are, would contain important information about the earthquake nucleation zone, as well. The above considerations are a strong motivation to study the magnetic field data set gathered around the $M_w$ 6.1 LAquila earthquake on April 6, 2009. The EQ occurred during very quiet geomagnetic activity, in night hours and in close vicinity to the magnetic observatory: 6.7 km to the epicenter and 10.7 km to the hypocenter – three very important circumstances to examine thoroughly all available data being of unprecedented quality.

Recently DiLorenzo et al (2011) have investigated magnetic field variations focusing on the main phase of the 2009 L'Aquila earthquake. The authors have analysed magnetic field data set sampled at higher frequency (10 Hz) measured at two magnetic stations: one installed close to the L'Aquila observatory (Palangio et al., 2009), located inside the seismogenic area and another one at Duronia (Karakelian et al., 2000) located outside the seismogenic region,

~130 km away from L'Aquila. The latter has been considered as a reference station. Based on differential measurements between the two permanent observatories of L'Aquila and Duronia, the residual magnetic field estimated by means of the inter-station impulse response functions in the frequency band from 0.3 Hz to 3 Hz has been examined (DiLorenzo et al, 2011). They have found the following results: 1) Clusters of ULF signals (of spectral content in the 0.3 Hz to 3 Hz range) were observed a) just a few minutes before and during the arrival of the first P seismic waves of the earthquake, and b) another burst was observed after the main phase (01:32:40.400 UT, 6 Apr. 2009), which lasted about 50 min; 2) The observed ULF signals were transmitted from the direction of the hypocenter. DiLorenzo et al (2011) have succeeded to reveal an existence of nearly co-seismic magnetic field disturbances belonging to the 0.3-3 Hz frequency window and most likely connected with processes in the hypocenter of the 2009 L'Aquila earthquake. The analysis of co-seismic magnetic field variations in ULF range 0.3-3 Hz was however disputed by Masci and De Luca (2013). They have reexamined the DiLorenzo et al's data and results and have concluded that there is no evidence to support the hypothesis that magnetic disturbances documented by Di Lorenzo et al. (2011) had a seismogenic origin.

Our goal is not to enter in contest, inversely we would draw an attention to the fact that DiLorenzo et al's analysis (2011) does not include the fluxgate magnetometers data set conducted there by L'Aquila University and INGV magnetic observatories and partly analyzed by Villante et al (2011) and Nenovski et al (2013). In these studies the co-seismic magnetic field variations in the ULF range were tentatively excluded and missed. The reason then was that the true signal, if exists, would be influenced by compromising mechanical and accidental displacements, undulations and vibrations appearing at the EQ shock moment impacting thus either some of or all the sensor axes.

In this study an attention will be paid to co-seismic magnetic field variations of longer temporal scale (of several tens of seconds and more) that would persist beyond the initial seismic shock effect. First, possible co-seismic travelling ionospheric disturbances possessing longer periods (minutes) could be expected and recognized by their magnetic field signatures. The latter are expected to appear several minutes later after the seismic wave train passage − the time delay is due to the fact that the gravity wave disturbances induced by seismic waves' Earth surface uplifting at the EQ shock need some time (say ~5 min) to reach the ionosphere where charge disturbances are initiated and the charge induced currents in the ionosphere induce thus travelling wavelike co-seismic magnetic signatures (for example, such ionosphere effects have been observed after the 1964 Alaska EQ). Thus, the main objective is an examination of expected co-seismic magnetic field variations occurred next to and at the 2009 Aquila mainshock.

Instead of wave-like signals the analysis, undertaken in this study, revealed an existence of transient magnetic field signal that appears immediately after the EQ shock. The signal exists within the first 5 minutes. The transient magnetic field signal emerging at the mainshock moment (01:32:40 UT) and lasting for several minutes after it, its temporal profile and polarization features are pinpointed in relation to the mainshock geometry (location) and dynamics.

*Geology and Geometry of the Aquila fault system.* The 2009 Mw 6.1 L'Aquila earthquake is one of the best-recorded normal faulting events concerning slip/displacement measurements and near-fault strong motion characteristics. The 2009 Mw 6.1 L'Aquila earthquake (red star, Figure 1) occurred in the Central Apennines (Italy) on 6 April at 01:32:40.400 UT, with the hypocenter located at 42.342° N, 13.38° E, and at a depth of 8.3 km (Orefice et al, 2013). The earthquake caused 300 and more casualties, and heavy damages in L'Aquila and several villages nearby. The EQ mainshock was preceded by a foreshock sequence starting a few months before (October 2008) and culminating with an $M_L$ 4.1 event on 30 March 2009, followed by $M_L$ 3.9 and $M_L$ 3.5 foreshocks on 5 April 2009 occurring a few hours before the mainshock. The earthquake rupture developed first up dip and then along northwest–southeast-striking segments of a complex normal fault system embedded in the mountain front of the central Apennines chain (Cirella et al., 2009; Walters et al., 2009; Orefice et al, 2013). Valoroso et al (2013) have investigated the complex fault architecture and mechanics associated with the 6 April 2009 Mw 6.1 L'Aquila earthquake, revealing that stresses were concentrated between high-angle normal faults that underwent unstable sliding (i.e. stick slip) and a basal low angle (<20 degrees) discontinuity at depths of ~9 km. According to Valoroso et al (2013) the L'Aquila fault breaks the entire upper crust from ~10 km depth to the surface, dipping 50 degrees (±2) to the SW, while its length is about 18 km along the N137E degrees direction. The Mw 6.1 main shock was nucleated at 8.27 km depth. Such a location of the main shock is slightly shallower compared with the one evaluated by Chiarabba et al. (2009) (equal to 9.46 km) and Chiaraluce et al. [2011] (equal to 8.64 km). Chiaraluce et al. (2011) have recognized two main SW-dipping normal faults, the L'Aquila and Campotosto segments, forming an en-echelon system, extending in the NW-SE direction for about 50 km. The L'Aquila fault itself is about 16 km long showing a nearly planar geometry with a constant dip (~48 degrees) between 10 and 2 km depth.

**Data**

The two magnetic observatories (belonging to INGV, Rome and L'Aquila University) are deployed in the Upper Aterno-river valley with coordinates 42.3826N, 13.316E and 42.381N, 13.3174E, respectively. The distance between the two observatories is appr. 200 m, i.e. in relation the L'Aquila earthquake mainshock the two observatories may be considered as one point measurement (blue triangle, Figure 1). Indeed, the mean distance to the EQ mainshock epicenter is 6.8 km. The calculated distance between the observatories and the EQ hypocenter (considered at depth 8.3 km) is 10.7 km. It is worth noting that accidentally the line connecting the magnetic observatories and the EQ hypocenter makes with the X axis (North) an angle of 48.7 degrees, i.e. this line is practically co-linear with the L'Aquila strike along the N137E degrees direction.

The INGV absolute magnetometers are fluxgate magnetometer for measurements of magnetic declination, D and inclination, I and an overhauser (absolute) magnetometer for measurement of total magnetic *F*. The fluxgate sensor possesses noise level ~15 pT/√Hz in the DC ÷1 Hz range (Palangio, pdf material). Overhauser magnetometers are unique in: i) keeping highest absolute accuracy of proton precession (this is primary standard for measurement of magnetic field in general); ii) improving greatly on sensitivities of proton magnetometers and enabling the highest accuracy to be practically achieved in weak magnetic fields such as the Earth's; and iii) allowing for continuous, uninterrupted measurement of the magnetic field of the Earth with sufficient speeds for any airborne work of study of fast phenomena occurring in the Earth's magnetic field (Hrvoic, 1989).

The L'Aquila University magnetic field observations available are those of the tri-axial fluxgate and induction magnetometers of the ULF station. Measurements from both instruments are recorded at a sampling frequency of 1 Hz from the same acquisition system. The fluxgate magnetometer has a rms instrumental noise of ~20 pT in the frequency band 1–500 mHz. According to Villante and Vellante (1998) and Villante et al. (2004), the man-made contamination determines in the ULF bands a weekly modulation with minimum power values on Sundays and reduced power level on Saturdays; such contamination would provide an additional noise during weekdays (daytime hours) of ~1 pT (200–500 mHz), ~2 pT (100–200 mHz), ~3 pT (20–100 mHz), while no significant contamination was determined in the lower frequency bands and, in general, during nighttime hours (Villante et al, 2010). For our analysis we exploit fluxgate and absolute (overhauser) magnetometer data exclusively on the earthquake day: April 06, 2009.

**Application and Results**

*A. The magnetic field offset*. The magnetic field offset is a well known phenomenon observed at earthquake shocks. Similar phenomenon was observed at the L'Aquila earthquake mainshock event. First high amplitude bursts of coseismic disturbances were recorded by the two fluxgate magnetometers. Their spectral contents however were quite different (correlation is absent) and as expected, they were presumably produced by the mechanical undulations and vibrations caused on the sensors by the EQ shock. The magnetic disturbances started simultaneously at all the magnetometers with a time delay/offset of ~5 seconds compared to the EQ mainshock moment fixed at 01:32:40.400 UTC (Orefice et al., 2013). Having in mind that the distance between the observatories and the EQ hypocenter is equal to 10.7 km, the delay time can be attributed to the travel time of the seismic waves to reach the measurement points where the shaking induces observed high amplitude magnetic field bursts. The estimated seismic wave velocity is thus estimated equal to 2.1 km/s. It thus lies in the lower part of possible seismic velocities range expected for the medium. Therefore, the magnetic field bursts were considered to be initiated simultaneously with the seismic waves that reached the observation point. Figure 2 displays the 10 second averaged magnetic field data illustrating the magnetic field offset event observed by fluxgate magnetometers. The offset magnitude as indicated on fluxgate magnetometer equipment was of different magnitudes for the three components:  ~ +0.5 nT on the x axis, ~ −1 nT − on the y axis, and of about −0.25 nT − on the z axis. The offset vector horizontally was directed mainly to EEN and vertically downwards.

The observed magnetic field offset parameters were consistent with that already discussed in literature. It was apparent that maximum signals are not more than a nanotesla or so and these signals occur only for larger earthquakes (M > 6) for which corresponding strain changes were about a microstrain or so (Johnson (2002). The piezomagnetic effect was the only mechanism involved many times for the explanation of such magnetic field slip events. The offset event, conventionally interpreted as a signature of piezomagnetic effects, however could not be ascribed as such. There is a substantial difference. The reason is that the presently known geology of the L'Aquila basin consists mainly of carbonates, dolomites and limestone. It does not thus suggest an appearance of piezomagnetized effects under EQ fracture/slip events. Instead, the observed offset marks an initial phase of a transient signal that evolves further.

*B. Transient magnetic field signal*. Figure 3 shows the variations of the geomagnetic field components recorded overhauser (absolute) magnetometer (belonging to INGV, Rome). The absolute magnetometer data revealed a transient magnetic field signal that initially starts as step-like jump reaching an amplitude maximum of 0.8 nT (Figure 3). The transient magnetic field signal grew after the seismic wave passage, retaining its maximum for about 15 seconds and begins to fade for about 200 seconds. The signal totally disappeared after 300 seconds (~5 minutes). The signal was local one and not dependent on the common geomagnetic activity. The overall geomagnetic activity on April 06, was quiet: the Kp index is well below 2; $D_{st}$ index is also extremely low (Villante et al, 2010; Nenovski et al, 2013, 2014). The geomagnetic field then reduces considerably their amplitude variations especially during the night hours. Figure 4 demonstrates the locality of the transient signal by comparing the 6 hours geomagnetic field variations on April 06, 2009 at two Italian stations L'Aquila (central Italy) and Castello Tesino (in North Italy).

The observed structure of the magnetic field signal was unique one – its amplitude profile was unusual when comparing to the background very quiet level of the total magnetic field variations occurred in the nighttime interval: 00:00-06:00 UT. If we 'exclude' both the short time burst effect (done by averaging) and the static offset (Figure 2), the structure of the variations in the $x$ and $y$ components is conspicuously the same, while the variations in the vertical $z$ component however seems undisturbed before, at and after the EQ shock. This finding was confirmed by subsequent spectral analysis. The dynamical spectra (frequency vs time) are displayed in Figure 5. It is seen the spectral distribution of the total field signal $F$ practically is consistent with the spectral distribution in $x$ and $y$ components recorded by the INGV fluxgate magnetometer. It is worth noting here that the second fluxgate magnetometer data (L'Aquila University) after the EQ shock moment revealed gross inconsistencies: the spectral distribution of the three components differs both qualitatively and quantitatively, i.e. the spectral distributions between any couple of components did not reveal any consistency. This was confirmed by an analysis of the data trends. Substracting the relevant fluxgate data sets, i.e. making data difference for each component, a stable device drift behavior (the difference amplitude grows with time) was revealed instead of a stable zero amplitude level. This fact was suggestive for a possible device malfunction. Later on, several hours after the EQ mainshock, the L'Aquila magnetometer recording finally broke. Hereafter, all the subsequent analysis is performed on the INGV magnetometer data.

The spectral analysis of the $x, y$ magnetic field components and the total field $F$ revealed an identical spectral composition and distribution in the studied diapason: 20-200 seconds (Figure 5). The peak amplitudes in $x$ and $y$ components were found to be nearly comparable; the two horizontal components quantitatively yielded a magnetic field disturbance $H$ of magnitude comparable to $F$. Again, the $z$ component did not possess transient signal properties (Figure 5) and practically did not contribute to the magnetic field signal revealed in the total magnetic field $F$. The transient magnetic field signal was thus lying practically in the horizontal plane, i.e. on the Earth surface. The polarization analysis yielded the following result: the magnetic vector of the observed signal was oriented in northwest direction crudely at 45 degrees from the $x$ (positive) axis. This was easily tested in a new frame of reference rotated counterclockwise at 45°. Remind that the L'Aquila fault strike is oriented in NW-SE direction, i.e. the horizontal vector of the transient signal coincided with the Aquila fault strike direction (137°). Correlation estimations between $x$, $y$ and $F$ data were in concordance with the spectral analysis.

Next we analyze temporal structure of the signal. The magnetic field signal may be divided in two parts: a short rising phase lasting for 15 seconds and less and a fading (relaxation) second phase having duration of 200 seconds and even more (Figure 6). It is worth noting that the initial moment of the rising part is masked by 'a step-like forerunner' of a 0.2÷0.3 nT coexisting with the offset event, probably produced by the seismic wave train passage, then the signal amplitude jumps suddenly (within one second) up to its maximum of 0.8 nT. After the maximum phase the signal resides asymptotically to zero.

The observed temporal structure of the observed magnetic field signal suggests a process of diffusion pattern:

$$F/F_0 \sim \frac{1}{\sqrt{t}} \exp(-x_0^2/4Dt),$$

where $F_0$ is a constant, $x_0$ - parameter characterizing some 1-D dimension, and $D$ – stands for a diffusion coefficient of the suggested diffusivity process; $t$ is time. Note that the assumed diffusion model is general one, in the sense that it is independent of any particular process. The question of whether such a model might be representative of actual physical processes accompanying EQ main shock however has to be left for a further consideration. At this moment we only pinpoint the universality of the diffusion process capable to explain the signal growth and fading. In particular, the assumed diffusion model might be directly testable because it allows a quantitative estimation of some physical parameters of the nucleation zone where the signal is expected to be generated. Hence, the suggested model might be confirmed or refuted by complementary observations. Figure 6 illustrates similarity between the experimental data and the assumed diffusion model contoured for three different diffusion time constants: $\tau \equiv x_0^2/4D$: 10, 5, 2.5 s. The second curve ($x_0^2/4D$ is equal to 5 s) qualitatively coincides with the experimental data. The result is only suggestive that the observed signal might be a manifestation of a diffusion process on the whole. Alternatively, if a relaxation process (of exponential form) was suggested, the characteristic time constant of the signal fading would amount to of about 70 s.

**Conclusion.** Co-seismic variations at the Mw6.1 EQ mainshock consist both in an offset and transient magnetic field signal of amplitudes ranging between 0.1-1 nT. They are recorded by simultaneous measurements with fluxgate and overhauser (absolute) magnetometers. The magnetic field observations around the L'Aquila earthquake mainshock reveal that i) a transient signal appears and fades within the first 5 minutes after the EQ shock and ii) it is indicative for an electrification process of diffusion type.

The second type of coseismic event, the transient magnetic signal starts simultaneously with the offset event and reaches amplitude of 0.8 nT in the total magnetic field. The signal is local one, in the sense that its form differs from the signals of ionospheric/magnetospheric origin confirmed (indirectly) by additional magnetic field data in Italy and Central Europe. The reliability of the observed local signal is examined also: it follows from the fact that the transient signal is recorded by two different measurements: i) in components (by fluxgate magnetometers) and 2) absolute one (overhauser magnetometer), it persists after the seismic wave train passage but within the first five minutes after the EQ shock and is not a consequence of ionposphere disturbances caused by seismic wave train. Its amplitude shape resembles diffusion-like form, its orientation is mainly in horizontal plane and nearly parallel to the L'Aquila fault strike, i.e. along the NW-SE direction.

The observed transient signal can be interpreted as an evidence of primary coseismic magnetic field signal presumably generated and/or diffused from the EQ nucleation zone.

**Acknowledgement** The authors acknowledge the magnetic field data quality provided by Dr Domenico Di Mauro, who is responsible for the database: http://roma2.rm.ingv.it/it/risorse/banche_dati/39/osservazioni_relative_al_sisma_del_6-4-2009_a_l-aquila,

*REFERENCES*

Chiarabba C., A. Amato, M. Anselmi, P. Baccheschi, I. Bianchi, M. Cattaneo, G. Cecere, L. Chiaraluce, M.G. Ciaccio, P. De Gori, G. De Luca, M. Di Bona, R. Di Stefano, L. Faenza, A. Govoni, L. Improta, F.P. Lucente, A. Marchetti, L. Margheriti, F. Mele, A. Michelini, G. Monachesi, M. Moretti, M. Pastori, N. Piana Agostinetti, D. Piccinini, P. Roselli, D. Seccia and L. Valoroso (2009), The 2009 L'Aquila (central Italy) MW6.3 earthquake: main shock and aftershocks, Geophys. Res. Lett., **36**, L18308, doi:10.1029/2009GL039627.

Chiarabba, C., S. Bagh, I. Bianchi, P. De Gori, and M. Barchi (2010), Deep structural heterogeneities and the tectonic evolution of the Abruzzi region (central Apennines, Italy) revealed by microseismicity, seismic tomography and teleseismic receiver functions, Earth Planet. Sci. Lett., 295, 462–476, doi:10.1016/j.epsl.2010.04.028.

Chiaraluce, L., L. Valoroso, D. Piccinini, R. Di Stefano, and P. De Gori (2011a), The natomy of the 2009 L'Aquila normal fault system (central Italy) imaged by high resolution foreshock and aftershock locations, J. Geophys. Res., 116, B12311, doi:10.1029/2011JB008352.

Cirella, A., A. Piatanesi, M. Cocco, E. Tinti, L. Scognamiglio, A. Michelini, A. Lomax, and E. Boschi (2009), Rupture history of the 2009 L'Aquila (Italy) earthquake from non-linear joint inversion of strong motion and GPS data, Geophys. Res. Lett., 36, L19304; doi:10.1029/2009GL039795.

Di Lorenzo, C., P. Palangio, G. Santarato, A. Meloni, U. Villante, and L. Santarelli, 2011. Non-inductive components of electromagnetic signals associated with L'Aquila earthquake sequences estimated by means of inter-station impulse response functions, Nat. Hazards Earth Syst. Sci., 11, 1047−1055, 2011;doi:10.5194/nhess-11-1047-2011.

Fenoglio, M.A., M.J.S. Johnson, and J.D. Byerlee (1995), Magnetic and electric field associated with changes in high pore pressure in fault zones: Application to the Loma Prieta ULF emissions, J. Geophys. Res., 100, No.B7, pp 12951-12958.

Hrvoic, Ivan, 1989. Overhauser Magnetometers For Measurement of the Earth's Magnetic Field, in *Magnetic field Workshop on Magnetic Observatory Instrumentation*, Espoo, Finland. 1989; see http://www.geophysik.uni-bremen.de/statisch/downloads/254/Overhauser-Magnetometers.pdf

Johnston, M. J. S., S. A. Silverman, R. J. Mueller, and K. S. Breckenridge, Secular variation, crustal contribution, and tectonic activity in California, 1976–1984, J. Geophys. Res.,90, 8707–8717, 1985.

Johnston, M. J. S., R. J. Mueller, and Y. Sasai, Magnetic field observations in the near-field of the 28 June 1992 Mw 7.3 Landers, California earthquake, Bull. Seism. Soc. Amer.,84, 792–798, 1994.

Johnston, M.J.S., 2002. Electromagnetic fields generated by earthquakes. International Handbook of Earthquake and Engineering Seismology, Volume 81A. New York: Academic Press, pp. 621−635.

Karakelian, D., S.L. Klemperer, A.C.Fraser-Smith G.C. and Beroza, 2000, A transportable system for monitoring ultra low frequency electromagnetic signals associated with earthquakes, Seism. Res. Lett., 71 423–436, 2000.

Masci, F. and G. De Luca, 2013, Some comments on the potential seismogenic origin of magnetic
disturbances observed by Di Lorenzo et al. (2011) close to the time of the 6 April 2009 L'Aquila earthquake, Nat. Hazards Earth Syst. Sci., 13, 1313–1319, 2013; doi:10.5194/nhess-13-1313-2013.

Nenovski, P., M. Chamati, U. Villante, M. De Lauretis, and P. Francia (2013) Scaling Characteristics of SEGMA Magnetic Field Data around the Mw 6.3 Aquila Earthquake, Acta Geophysica, 61, No. 2, 2013, pp. 311-337; doi: 10.2478/s11600-012-0081-1.

P.I. Nenovski, M. Pezzopane, L. Ciraolo, M. Vellante, U. Villante, M. De Lauretis, 2014. Local changes in the total electron content immediately before the 2009 Abruzzo earthquake, Adv. Space Res., 55, 1, 2015, pp 243–258; http://dx.doi.org/10.1016/j.asr.2014.09.029

Orefice, A., M. Vallée, J. Balestra, B. Delouis, and A. Zollo, 2013, Refined Rupture-Velocity Estimation of the 2009 L'Aquila Earthquake (Mw 6.3, Central Italy) Derived from Apparent Source Time Functions, Bulletin of the Seismological Society of America, Vol. 103, No. 4, pp. 2474–2481, August 2013, doi: 10.1785/0120120255.

Palangio, P., 2009. Magnetism and Electromagnetismin Central Italy, INGV -L'Aquila Geomagnetic Observatory, a pdf material.

Simpson, J. J., and A. Taflove (2005), Electrokinetic effect of the Loma Prieta earthquake calculated by an entire-Earth FDTD solution of Maxwell's equations, Geophys. Res. Lett., 32, L09302, doi:10.1029/2005GL022601

Valoroso, L., L. Chiaraluce, D. Piccinini, R. Di Stefano, D. Schaff, and F. Waldhauser (2013), Radiography of a normal fault system by 64,000 high-precision earthquake locations: The 2009 L'Aquila (central Italy) case study, J. Geophys. Res. Solid Earth, 118, doi:10.1002/jgrb.50130.

Villante U. and M. Vellante, 1998, An analysis of working days contamination in the icropulsation band, Ann. Geofis., 48, 325−332.

Villante, U., M. Vellante, A. Piancatelli, A. Di Cienzo, T.L. Zhang, W. Magnes, V. Wesztergom, V., and A. Meloni, 2004, Some aspects of man-made contamination on ULF measurements, Ann. Geophys., 22, 1335–1345, 2004, http://www.ann-geophys.net/22/1335/2004/.

Villante, U., M. De Lauretis, C. De Paulis, P. Francia, A. Piancatelli, E. Pietropaolo, M. Vellante, A. Meloni, P. Palangio, K. Schwingenschuh, G. Prattes, W. Magnes, and P. Nenovski, 2010. The 6 April 2009 earthquake at L'Aquila: a preliminary analysis of magnetic field measurements, Nat. Hazards Earth Syst. Sci., 10, 203–214, 2010.

Walters, R. J., J. R. Elliott, N. D 'Agostino, P. C. England, I. Hunstad, J. A. Jackson, B. Parsons, R. J. Phillips, and G. Roberts (2009). The 2009 L'Aquila earthquake (central Italy):

A source mechanism and implications for seismic hazard, Geophys. Res. Lett. 36, L17312; doi: 10.1029/2009GL039337
.

**FIGURE CAPTIONS**

*Figure 1*. USGS Shaking Map of the L'Aquila earthquake epicenter in central Italy. Red star marks the earthquake epicencer and the blue triangle - the magnetic observatories in North-West from L'Aquila.

*Figure 2*. The 10 seconds average of fluxgate and overhauser magnetic field data set around the L'Aquila earthquake (EQ) main shock occurred on 01:32:40.400 UT. On the first two panels (x and y components), the magnetic offset is clearly observed. On the fourth panel a transient magnetic field signal appears at the EQ mainshock and disappears within the first 5 minutes after the shock moment.

*Figure 3*. The transient magnetic field signal. Details of its temporal structure.

*Figure 4*. Geomagnetic field variations of two hours length recorded at two points: the L'Aquila observatory and Castello Tesino station (in North Italy). The transient signal appears at L'Aquila only. It differs from the common variations produced by geomagnetic activity (very quiet for the period).

*Figure 5*. Dynamic spectra of the magnetic field data set. The period range starts at 20 seconds. The induced vibrations by seismic waves appear mainly at periods below 10 seconds and thus they practically are eliminated.

*Figure 6*. The transient magnetic field signal and diffusion model with three time constants are displayed. All peak amplitudes are normalized to 1.

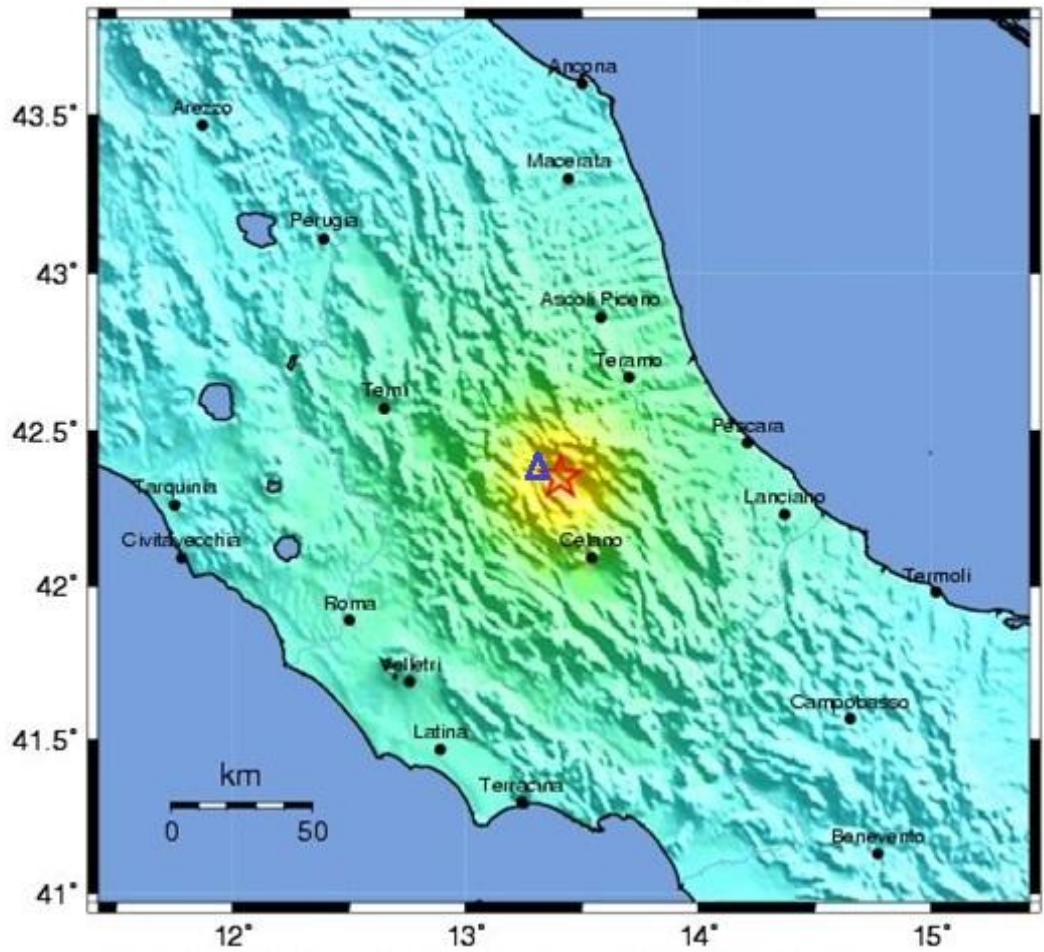

**Figure 1**

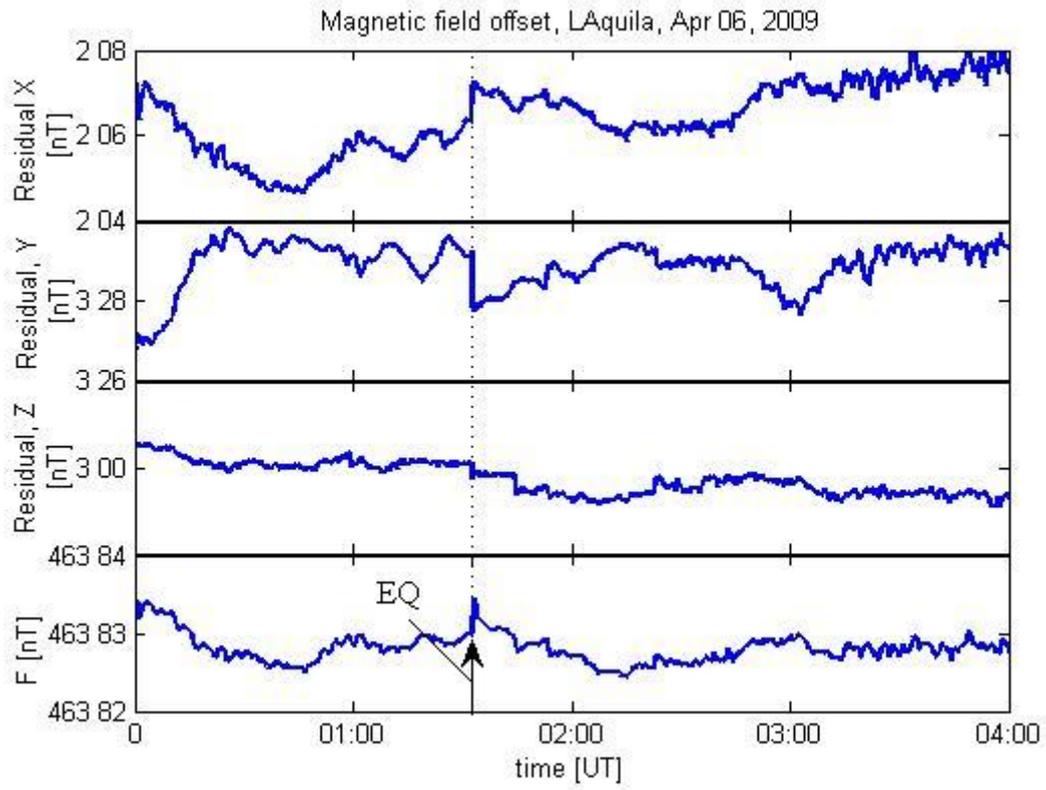

**Figure 2**

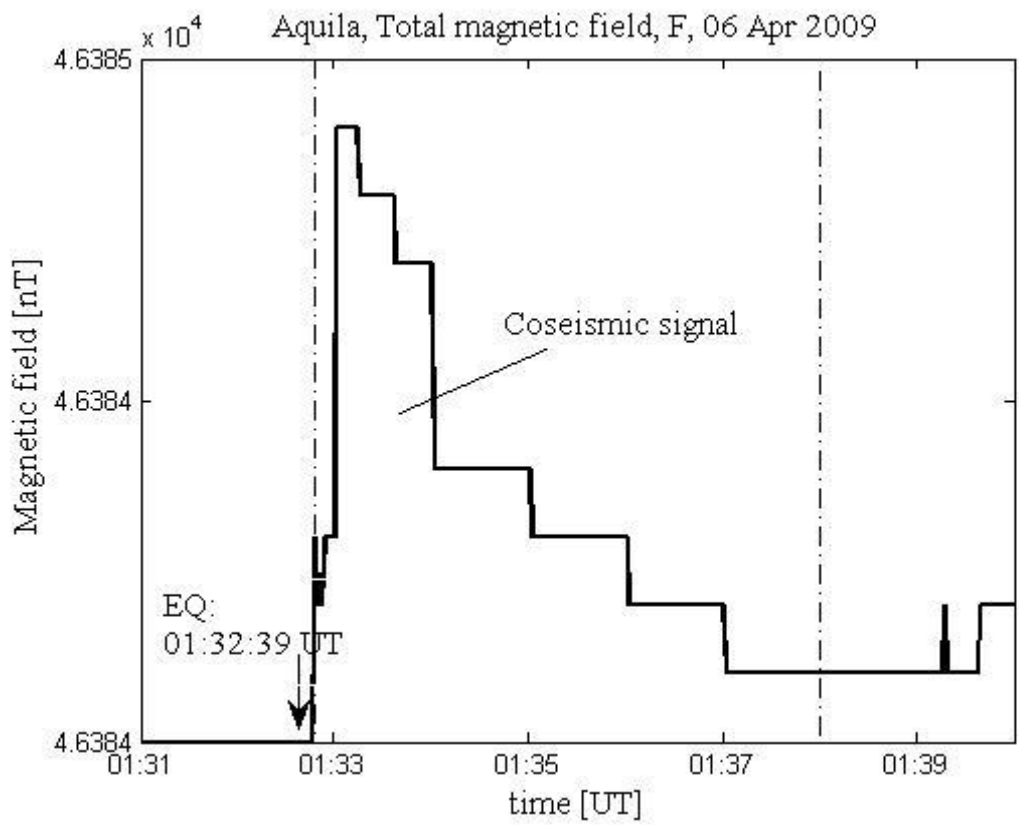

**Figure 3**

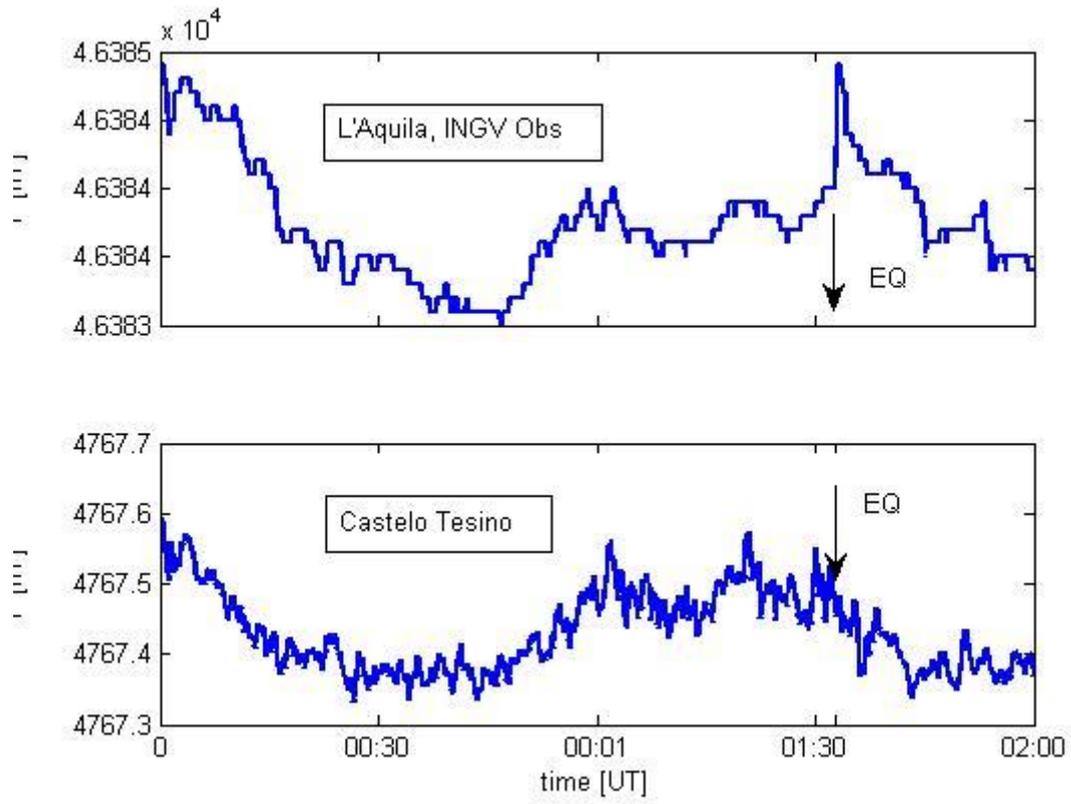

**Figure 4**

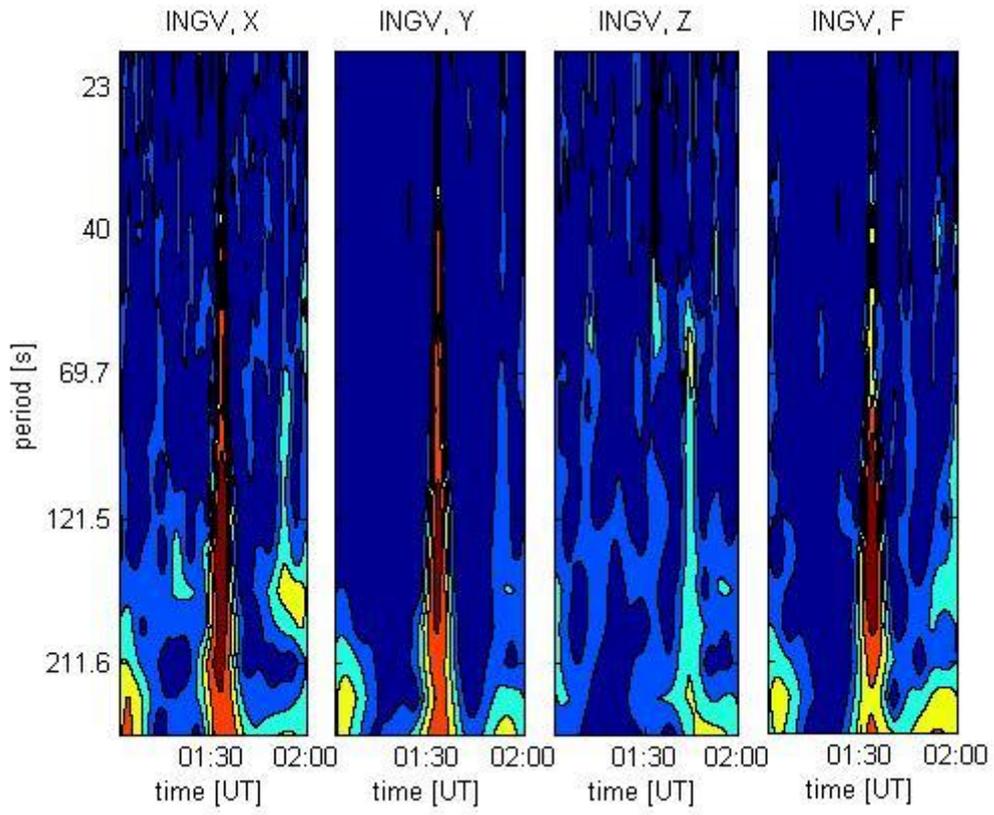

**Figure 5**

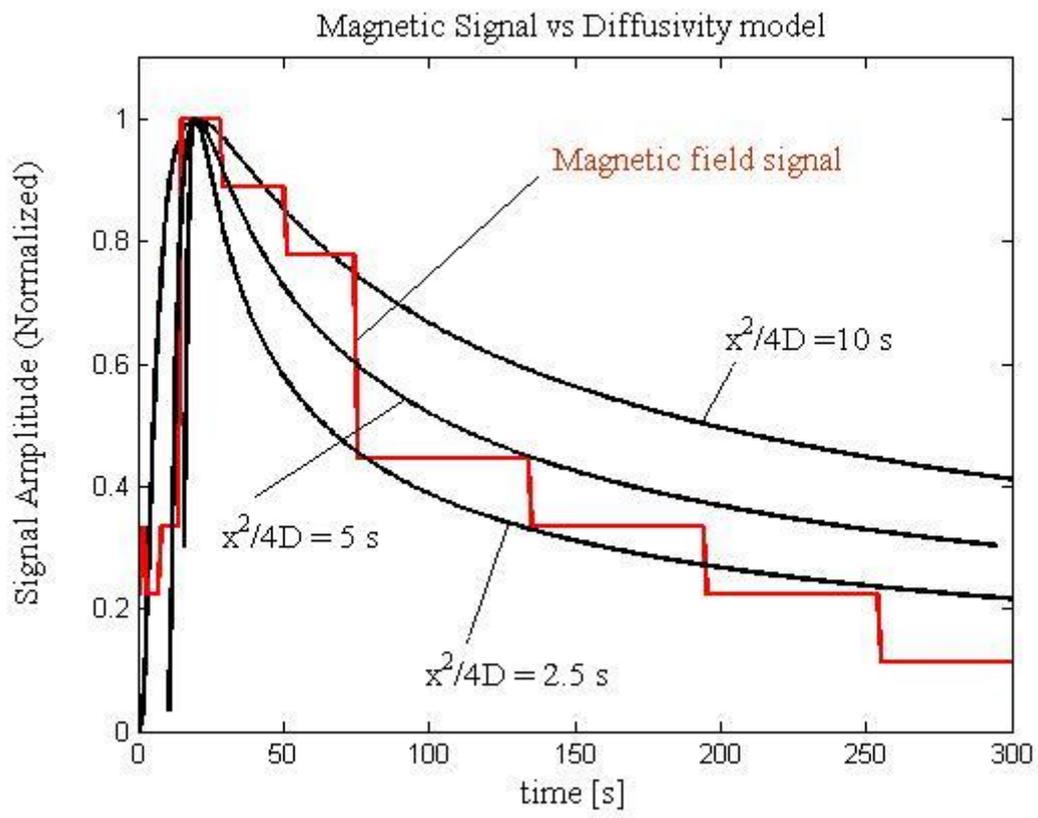

**Figure 6**